\documentclass{JHEP3}

\title{Oscillations of neutrino velocity}
\author{Branislav Sazdovi\'c, Milovan Vasili\'c  \\
Institute of Physics, University of Belgrade, P.O.Box 57, 11001 Belgrade, Serbia \\
E-mail: \email{sazdovic@ipb.ac.rs}, \email{mvasilic@ipb.ac.rs} }

\abstract{ In this paper, we consider the problem of quantum
measurement of neutrino velocity. We show, that the well known
neutrino flavor oscillations are always accompanied by the
oscillations of neutrino velocity. In particular, the velocity
of a freely moving neutrino is demonstrated to periodically
exceed the speed of light. Unfortunately, the superluminal
effect turns out to be too small to be experimentally detected.
It is also shown that neutrino velocity significantly depends on
the energy, size and shape of the neutrino wave packet. Owing to
the big experimental error of the recent experiments, these
dependences remained unnoticeable. Finally, we have shown that
the recent claims that superluminal neutrinos should loose
energy during their flight is not true. Instead, our formula
suggests the approximate conservation of energy along neutrino
trajectory. All these results have been obtained without
violation of special theory of relativity. }


\preprint{}

\usepackage{amsmath,amssymb,epsfig}

\newcommand{\q}{\hphantom{s}}
\newcommand{\s}{\hphantom{ss}}
\newcommand{\btau}{\mbox{\boldmath $\tau$}}

\newcommand{\ds}{\displaystyle}

\begin{document}

\section{\label{introduction}Introduction}

The existence of neutrino flavor oscillations is considered a well
established fact in contemporary physics \cite{01,02,03,04,05,06,07,08,09}.
It is caused by the fact that the three known neutrino flavors, $\nu_{e}$,
$\nu_{\mu}$ and $\nu_{\tau}$, are not the eigenstates of the Hamiltonian.
Instead, they are superpositions of the true eigenstates $\nu_1$, $\nu_2$
and $\nu_3$, having sharp masses $m_1$, $m_2$ and $m_3$, respectively. As a
consequence, the flavors $\nu_{e}$, $\nu_{\mu}$ and $\nu_{\tau}$ oscillate
during the time evolution of a free neutrino.

The purpose of this work is to examine if similar oscillatory character may
appear when it comes to neutrino velocity. It is motivated by the observation
that flavor oscillations necessarily imply the oscillations of neutrino
masses. Then, owing to the momentum conservation, we expect a
freely moving neutrino to have oscillating speed.

The idea that neutrino velocity is closely related to the flavor
oscillations is not new. It has been explored in refs \cite{2,3,4}, with the
result that neutrinos are superluminal. However, our analysis
shows that the main expression of refs \cite{2,3,4} represents
just a small correction to our result. In particular, we demonstrate that
neutrino velocity considerably depends on the size and shape of the
neutrino wave packet.

In this paper, we shall work in the approximation of just two
flavors: $\nu_{\mu}$ and $\nu_{\tau}$. We shall demonstrate that
the velocity of the free muon neutrino indeed has oscillating
character. In particular, the neutrino velocity periodically
exceeds the speed of light. The maximal value of the detected
velocity along neutrino trajectory has also been found.
Interestingly enough, the probability to detect maximal velocity
is pretty small, and in some cases, goes to zero. At the same
time, the obtained formulae turn out to be unexpectedly
sensitive to the shape and size of the neutrino wave packet.
This makes the comparison with the known experiments very
difficult. For one thing, the exact shape of the experimental
wave packets is not known. For the other, we are not convinced
that all the packets in the ensemble are identical.
Nevertheless, we have tested our formula by comparing its
predictions with three recent experiments \cite{a,b,1}. For that
purpose, the numerical values of our free parameters (such as
neutrino energy) are chosen from these experiments, and the
undetermined free parameters, related to the wave packet shape
and size, were chosen by consulting the literature \cite{c}. As a result,
a good agreement with related measurements has been achieved. In
particular, the derived energy dependence of the neutrino
velocity ($v_{eff}-1\, \sim\, 1/E^4\, {\rm or}\, 1/E^6$) has
been shown to remain undetectable in the considered experiments.
The same holds for the apparent loss of energy during the flight
of superluminal neutrinos \cite{d}. Indeed, we have shown that
the rate at which superluminal neutrinos loose their energy is
linear in time, but the slope of the graph $E(t)$ is extremely
small. This way, the loss of neutrino energy becomes
unnoticeable in all terrestrial experiments.

In what follows, we shall use the natural units $\hbar = c = 1$.

\section{\label{dynamics}Neutrino dynamics}

To simplify the study of neutrino oscillations, in what follows, we shall
adopt two useful approximations. The first is that only two flavors,
$\nu_{\mu}$ and $\nu_{\tau}$, will be considered to span the internal
Hilbert space of the neutrino (the remaining internal degrees of freedom
will be neglected). The second is that the neutrino moves along the
$x$-axes, and its dependence on $y$ and $z$ is considered irrelevant. In
practice, this means that we reduce our task to a one-dimensional problem.
With this in mind, the Hilbert space of the neutrino becomes ${\cal H} = C^2
\otimes {\cal H}_0$, where $C^2$ is the internal Hilbert space spanned by
two orthonormal vectors $| i \rangle$, ($i=1,2$), and ${\cal H}_0$ is
orbital Hilbert space spanned by the momentum eigenvectors $| p \rangle$,
($-\infty < p < \infty$). The basis vectors $| i \rangle | p \rangle$ are
taken to be the eigenvectors of the Hamiltonian of the free neutrino:
\begin{equation}\label{1}
\hat H \, | i \rangle | p \rangle = E_i \, | i \rangle | p \rangle \,,
\qquad  E_i \equiv \sqrt{m^2_i + p^2} \,.
\end{equation}
Thus, the states $| i \rangle | p \rangle$ have the sharp values of mass.
On the contrary, the muon and tau neutrinos are not the eigenstates of the
Hamiltonian. Precisely, the general $\nu_{\mu}$ and $\nu_{\tau}$ states
have the form
\begin{eqnarray}
| \nu_{\mu} \rangle &=& \int dp\,a(p) \big(
\cos\theta\,|1\rangle |p\rangle -
\sin\theta\, | 2 \rangle | p \rangle \big) \label{2} \,, \\
| \nu_{\tau} \rangle &=& \int dp\,b(p) \big(
\sin\theta\,|1\rangle |p\rangle +
\cos\theta\, | 2 \rangle | p \rangle \big) \label{3} \,,
\end{eqnarray}
where $\theta$ is the mixing angle determined by $0.92 \lesssim \sin^2
2\theta \lesssim 1$, and
$$
\int dp\, |a(p)|^2 = \int dp\, |b(p)|^2 = 1
$$
ensures the proper normalization.

In what follows, we shall consider the temporal evolution of {\it initially pure
muon neutrino}. Its generic state vector is given by (\ref{2}), and its
evolution is determined by the Hamiltonian (\ref{1}). Thus, we obtain
$$
| \nu_{\mu}(t)\rangle =
\int dp\,a(p) \Big( \cos\theta\, e^{-iE_1 t}\,| 1 \rangle | p \rangle -
\sin\theta\, e^{-iE_2 t}\,| 2 \rangle | p \rangle \Big)  .
$$
The probability density to detect the muon neutrino in point $x$ is then
given by
$$
P_{\mu}(x,t) = \left| \langle \nu_{\mu}(x) | \nu_{\mu}(t)\rangle
\right|^2 \,,
$$
where $|\nu_{\mu}(x)\rangle \equiv \cos\theta\, | 1 \rangle | x \rangle -
\sin\theta\, | 2 \rangle | x \rangle$ stands for the eigenstate of the
position operator for muon neutrino. Direct calculation then yields
\begin{equation}\label{8}
P_{\mu}(x,t) = \frac{1}{2\pi} \left|
\int dp\,a(p) e^{ipx} \Big( \cos^2\theta\, e^{-iE_1 t} +
\sin^2\theta\, e^{-iE_2 t} \Big)  \right|^2 .
\end{equation}
To solve this integral, we shall use the reasonable assumption that the
distribution $a(p)$ is sharply localized around the value $p=p_0$. This way,
the neutrino is characterized by an almost sharp value of momentum. Then, we
can expand the energy $E_i(p) \equiv \sqrt{m_i^2 + p^2}$ in a power series
around $p=p_0$, and keep only linear terms. We shall also introduce the
group velocities
$$
v_i \equiv \frac{dE_{i0}}{dp_0} \,,
$$
and the shorthand notation $E_{i0} \equiv E_i(p_0)$. With these
approximations, the integration in (\ref{8}) leads to
\begin{equation}\label{11}
P_{\mu}(x,t) = \left|  A_1 \cos^2\theta\, e^{-i (E_{10}
-v_1 p_0 ) t} + A_2 \sin^2\theta\, e^{-i (E_{20}-v_2 p_0 )
t} \right|^2 ,
\end{equation}
where $A_i \equiv A(v_i t - x)$, and
$$
A(\tau) \equiv \frac{1}{\sqrt{2\pi}} \int dp \, a(p) e^{-ip\tau}  \,.
$$
The amplitude $A(\tau)$ in the coordinate space is the exact Fourier transform
of the momentum amplitude $a(p)$.  We have already assumed that the momentum
of the neutrino wave packet is well localized around $p=p_0$. To take this
explicitly into account, we shall take the neutrino wave packet in the form
\begin{equation}\label{12a}
A(\tau) = \rho(\tau) e^{-i \tau p_0}  \,  ,
\end{equation}
where the modulus $\rho(\tau)$ is localized around $\tau = 0$,
and the wavelength $2\pi/p_0$ is much smaller than the packet
size. The latter ensures a small uncertainty of the packet
momentum. In what follows, we shall see how our results depend on
the size and shape of the neutrino wave packet.

The values of energy and velocity of the two terms in (\ref{11})
differ as a consequence of different masses they carry. To estimate
their difference, we introduce
$$
\omega \equiv \frac{E_{20} - E_{10}}{2} \,,\qquad
\Delta v \equiv v_2 - v_1   \,, \qquad
\Delta m^2 \equiv m^2_2 - m^2_1  \,.
$$
In the ultrarelativistic limit $p_0 \gg m_i$, suitable for the description
of neutrinos, $E_{i0}$ is further decomposed as
\begin{equation}\label{ei}
E_{i0} = p_0 + \frac{m_i^2}{2p_0} + \cdots  \quad \Rightarrow \quad
\omega = \frac{\Delta m^2}{4p_0} + \cdots \,,
\end{equation}
and the velocities $v_i$ take the form
\begin{equation}\label{vi}
v_i =  1 - \frac{m_i^2}{2p_0^2} + \cdots  \quad \Rightarrow \quad
\Delta v = - \frac{2 \omega}{p_0} + \cdots  \,.
\end{equation}
Let us now analyze the probability density (\ref{11}) in more
detail. First, observe that the velocities $v_1$ and $v_2$,
although close to the speed of light, are different from each
other. Owing to this, the two initially overlapping packets will
gradually separate. After a long enough time, we shall see two
distinct neutrino wave packets. To estimate the time needed for
the separation of the two packets, we make use of the packet
size. Then, the time needed for their minimal separation (when
the distance between the packets reaches the packet size) is
given by
$$
t = \frac{2\ell}{|\Delta v|} = \frac{4\ell p_0^2}{\Delta m^2}  \,,
$$
where $\ell$ is half the size of the wave packet. Using the
numerical data from the recent experiments \cite{a,b,1} (as
shown in Table \ref{t2} of the last section), we find
$$
t\, \gtrsim\, 5 \, {\rm min} ,
$$
telling us that the two wave packets in (\ref{11}) practically coincide in
any terrestrial experiment ($5$ min $\approx$ $10^8$ km). For this reason,
in what follows, we shall simplify our considerations by adopting the restriction
$$
t|\Delta v| \ll \ell \,.
$$
Note that this restriction still allows for very long flights ($x \sim 10^6$ km).

Let us evaluate the amplitudes in the first order in the small parameter
$t\Delta v$. To this end, we introduce a new time coordinate $\tau$, and the
average velocity $\bar v$, defined by
\begin{equation}\label{15}
t \equiv \frac{x}{\bar v} + \tau \,, \qquad
\bar v \equiv \frac{v_1 + v_2}{2} \,.
\end{equation}
The new time coordinate measures time relative to the moment a particle
with the average velocity $\bar v$ arrives at $x$. Having in mind that
$\bar v \approx 1$, we can say that {\it $\tau$ measures the neutrino
delay as compared to the arrival time of the photon}. The
moduli of the amplitudes now become:
\begin{eqnarray}
&&  \rho(v_1\, t - x) =
\rho(\bar v\tau - \frac{\Delta v}{2} t) \approx
\rho (\bar v\tau )-\frac{t\Delta v}{2} \rho^\prime (\bar v \tau) \,,  \nonumber  \\
&&  \rho(v_2\, t - x) =
\rho(\bar v\tau + \frac{\Delta v}{2} t) \approx
\rho (\bar v\tau )+\frac{t\Delta v}{2} \rho^\prime (\bar v \tau) \,.  \nonumber
\end{eqnarray}
Owing to the smallness of the factor $\left|\frac{\Delta v}{2}\right|$,
the second term is expected to be much smaller than the first one.
Indeed, the numerical value of this factor in the recent experiments
\cite{a,b,1} is less than $10^{-22}$. At the same time, the phase of
$A$, as defined in (\ref{12a}), is subject to no approximation at all.

With the adopted approximations, the probability density (\ref{11}) takes
the form
\begin{equation}\label{17d}
P_{\mu}(x,t) = \rho^2 (\bar v\tau)\left( 1 - \sin^2 2\theta\,
\sin^2 \omega t \right) - [\rho^2(\bar v \tau)]^\prime  \cos2\theta \,
\frac{\Delta v}{2}t \,,
\end{equation}
where $\bar v\tau \equiv \bar v t-x$, and the prime denotes derivative with
respect to the argument. As the realistic wave packets have finite size, it
is natural to assume that the amplitude $\rho(\tau)$ is localized in the
interval $-\ell < \tau < \ell$. This way, the time coordinate $\tau$ is
restricted by the packet size, which is typically much smaller than $x$. In
what follows, we shall adopt the reasonable restriction
\begin{equation}\label{16b}
 |\tau| < \ell \ll  x \,.
\end{equation}

\section{\label{velocity}Velocity oscillations}

In this section, we shall study the motion of muon neutrino by studying the
spacetime dependence of its probability distribution (\ref{17d}). To this
end, let us first place the neutrino detector in a fixed position $x$. This
way, the probability to detect the muon neutrino becomes a function of time,
only. The {\it moment neutrino arrives at the detector is determined as the
time the probability density (\ref{17d}) reaches its maximum in the point} $x$. The
needed time of arrival is then obtained by solving the equation
$$
\frac{\partial P_{\mu}}{\partial t} =0 \,.
$$
With the approximation $\bar v\approx 1$, it ultimately leads to
\begin{equation}\label{19}
\frac{\partial_{\tau}\rho^2}{\rho^2}
-\frac{\partial^2_{\tau}\rho^2}{\rho^2} \btau_2 (t) +
\frac{4}{\ell^2} \btau_1 (t) =0   \, ,
\end{equation}
with
\begin{equation}\label{tau1}
\btau_1 (t) \equiv -\, \frac{\omega \ell^2}{4 }\, \frac{\sin^2
2\theta\,\sin 2\omega t} {1- \sin^2 2\theta\,\sin^2  \omega t  +
\frac{\omega}{p_0}\cos 2\theta}  \, ,
\end{equation}
\begin{equation}\label{tau2}
\btau_2 (t) \equiv  - \frac{\omega}{p_0}  \frac{\cos 2\theta}
{1- \sin^2 2\theta\,\sin^2 \omega t  + \frac{\omega}{p_0}\cos
2\theta} \,\, t \, .
\end{equation}
Before we continue, note that the restriction (\ref{16b}) allows us to
easily switch between the $t$ and $x$ coordinates. Indeed, after taking
into account $|\tau| \ll x$, and $\bar v \approx 1$, the equation
(\ref{15}) shows that
$$
 t \approx  x \,.
$$
With this, every solution of (\ref{19}), which is originally in the form
$\tau = \tau (t)$, can approximately be rewritten in the form $\tau =
\tau(x)$.

\subsection{Example of a simple wave packet}\label{A}

The above analysis is the most we can do without specifying the amplitude
$A$. As it turns out, the solution $\tau = \tau(x)$ depends a great deal on
the size and shape of the wave packet. Before we carefully examine this
dependence in the next section, let us describe the simple example of a
particular wave packet. Up to the normalization constant, we define the
modulus of the amplitude (\ref{12a}) as
\begin{equation}\label{23}
\rho (\tau) \propto \left\{
\begin{array}{cc}
\ds 1- \frac{\tau^2}{\ell^2}\,, & -\ell < \tau < \ell \\
0 \,, & \ \ \,{\rm otherwise}\ ,
\end{array} \right.
\end{equation}
where the parameter $\ell$ determines the packet size.
To simplify calculations, we shall work in the approximation $\tau \ll
\ell$. First, we multiply the equation (\ref{19}) with $\ell^2-\tau^2$, and
expand it in a power series of $\tau/\ell$. Then, we drop terms
proportional to $\tau^4/\ell^4$, whereupon the equation (\ref{19}) becomes
the quadratic equation
\begin{equation}\label{45b}
(\btau_1 + 2\btau_2) \,\tau^2 + \ell^2\,\tau
-(\btau_1 + \btau_2) \ell^2 = 0 \, .
\end{equation}
As seen from (\ref{45b}), $\btau_i \to 0$ implies $\tau\to 0$,
which uniquely determines the physical solution of the quadratic
equation to be
\begin{equation}\label{47c}
\tau (x) = -\frac{ \ell^2}{2 (\btau_1 + 2\btau_2)}
\left(1- \sqrt{1 + 4 \frac{ \btau_1 + \btau_2 }{\ell}
\cdot \frac{ \btau_1 + 2\btau_2 }{\ell} }  \right)  \,.
\end{equation}
On the other hand,  if we restrict our considerations to the linear
approximation in $\tau/\ell$, the equation (\ref{45b}) yields
\begin{equation}\label{22z}
\tau (x)= \btau_1 (x) + \btau_2 (x) \,.
\end{equation}
What is immediately seen is that $\btau_1 (x)$  {\it is a
periodic function of $x$} with the period
\begin{equation}\label{22}
L = \frac{\pi}{ \omega} \,.
\end{equation}
(For illustration, the neutrino energy of $17$ GeV yields $L \approx 18,000
\, {\rm km}$). The term $\btau_2 (x)$  is linear in $x$, and the
coefficient in front of it is periodic with the same period $L$.
In fact, this term is the main result of Refs. \cite{2,3,4}. It
does not depend either on the size or on the shape of the wave
packet. Owing to the linear dependence on $x$, the term
$\btau_2$ can exceed $\btau_1$ in the limit $x \to \infty$.
Let us accurately estimate the conditions needed to prevent
$\btau_2$ from dominating $\btau_1$. Thus, we
start with $|\btau_2| \ll |\btau_1|$, and obtain
\begin{equation}\label{22c}
\frac{\pi}{4} - \theta \ll \frac{p_0\ell^2}{8}\,\frac{|\sin 2\omega x|}{x} \,.
\end{equation}
Generally, this condition can not be fulfilled for all
distances. However, if we avoid distances which are too close to
$nL/2$, the requirement (\ref{22c}) can be solved for $x$. For
demonstration purposes, let us take data from the recent
experiments \cite{a,b,1}, as shown in Table \ref{t2}. Then, one
finds $p_0\ell^2/8 \gtrsim 1.15 \cdot 10^8$ km, and
$\frac{\pi}{4}-\theta \lesssim 0.14$, which, in the worst
scenario, yields $x \ll 10^9\, {\rm km}$. So, if $x$ is not too
close to $nL/2$, we can neglect $\btau_2$ for all
terrestrial distances. With this in mind, the formula
(\ref{47c}) becomes
\begin{equation}\label{477}
\tau (x) = -\frac{ \ell^2}{2\,\btau_1 }\left(1- \sqrt{1 +
\left(\frac{ 2 \,\btau_1}{\ell}\right)^2 }    \right)  \,,
\end{equation}
while linear approximation yields
\begin{equation}\label{24x}
\tau (x)= \btau_1 (x) \,.
\end{equation}
In addition, $\btau_1$ can be rewritten without the term proportional to the
extremely small coefficient $\omega/p_0$. Indeed, the numerical value
of this factor in the recent experiments \cite{a,b,1} is less than $10^{-22}$.
Thus, with the big precision, we can write
\begin{equation}\label{24y}
\btau_1 \approx -\,\frac{\omega \ell^2}{4}\,
\frac{\sin^2 2 \theta\,\sin 2\omega x} {1- \sin^2 2\theta\,\sin^2
\omega x }  \,.
\end{equation}
%

\subsection{Graphic illustration}\label{B}

Let us now analyze the consequences of the formula (\ref{477}). As we
can see, the neutrino delay $\tau(x)$ takes negative values when $x
\in [n L , (n+\frac{1}{2})L]$, positive values when $x \in
[(n+\frac{1}{2})L , (n+1)L]$ and zeros when $x= \frac{n}{2}L$ for all
integers $n\geq 0$. For demonstration purposes, let us make a numerical
example by choosing our free parameters as follows:
\begin{equation}\label{example}
p_0 \approx 17\ {\rm GeV}\,, \qquad
\ell \approx 1.4\ {\rm  km}.
\end{equation}
With the known values of $\Delta m^2 \approx 2.3\cdot 10^{-3}\, eV^2$, and
$0.92 \lesssim \sin^2 2\theta \lesssim 1$, this yields the behaviour shown
in Fig. \ref{fg1}.
%
\FIGURE{
\hbox{\epsfysize=7.2cm\epsffile{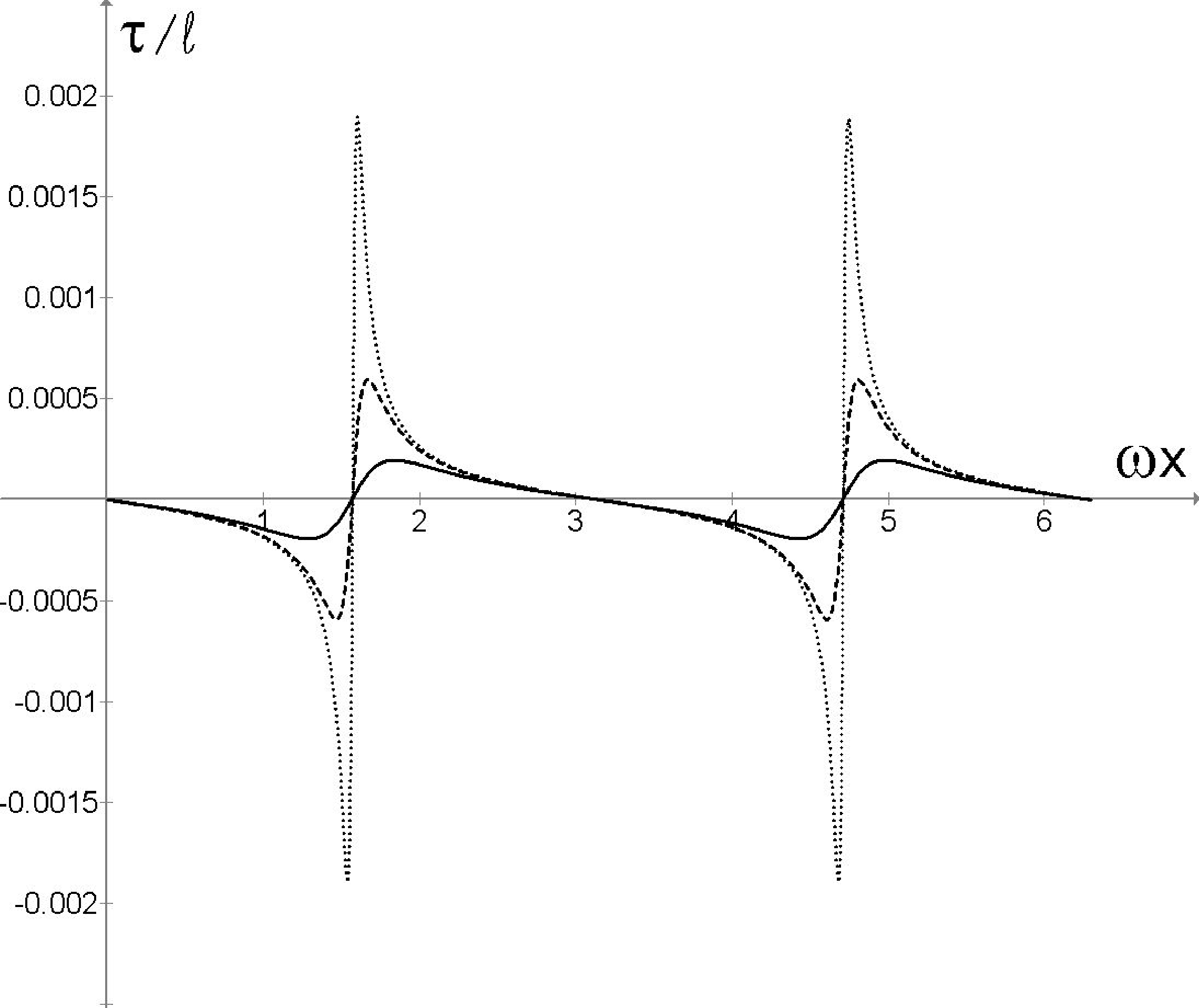}}
\caption{Oscillations of neutrino delay.
The three cases are defined by $\sin^2 2\theta = 0.92$
(solid line), $\sin^2 2\theta = 0.99$ (dashed line), and $\sin^2 2\theta = 0.999$
(dotted line).} \label{fg1}
}
%
The negative values of the neutrino delay time $\tau$ indicate that
neutrino may arrive earlier than expected on the basis of the average
velocity $\bar v$. Having in mind that $\bar v \approx 1$, we expect
that the neutrino velocity
$$
v(x) \equiv \frac{dx}{dt} \approx 1- \frac{d\tau(x)}{dx} \,  ,
$$
may periodically exceed the speed of light. Indeed, differentiating
(\ref{477}), we obtain
$$
v(x) =1+ 4 \frac{\sqrt{1 + \left(\frac{ 2 \,\btau_1}{\ell}\right)^2 }-1}{\sqrt{1 +
\left(\frac{ 2 \,\btau_1}{\ell}\right)^2 }}\,\,\frac{1-(1+\cos^2 2 \theta)\sin^2 \omega x}
{\sin^2 2\theta\,\sin^2 2\omega x }  \,.
$$
As expected, the neutrino velocity $v(x)$ is a periodic function of $x$,
with the period $L$. Its behavior is shown in Fig. \ref{fg2}. As we can
%
\FIGURE{
\hbox{\epsfysize=7.2cm\epsffile{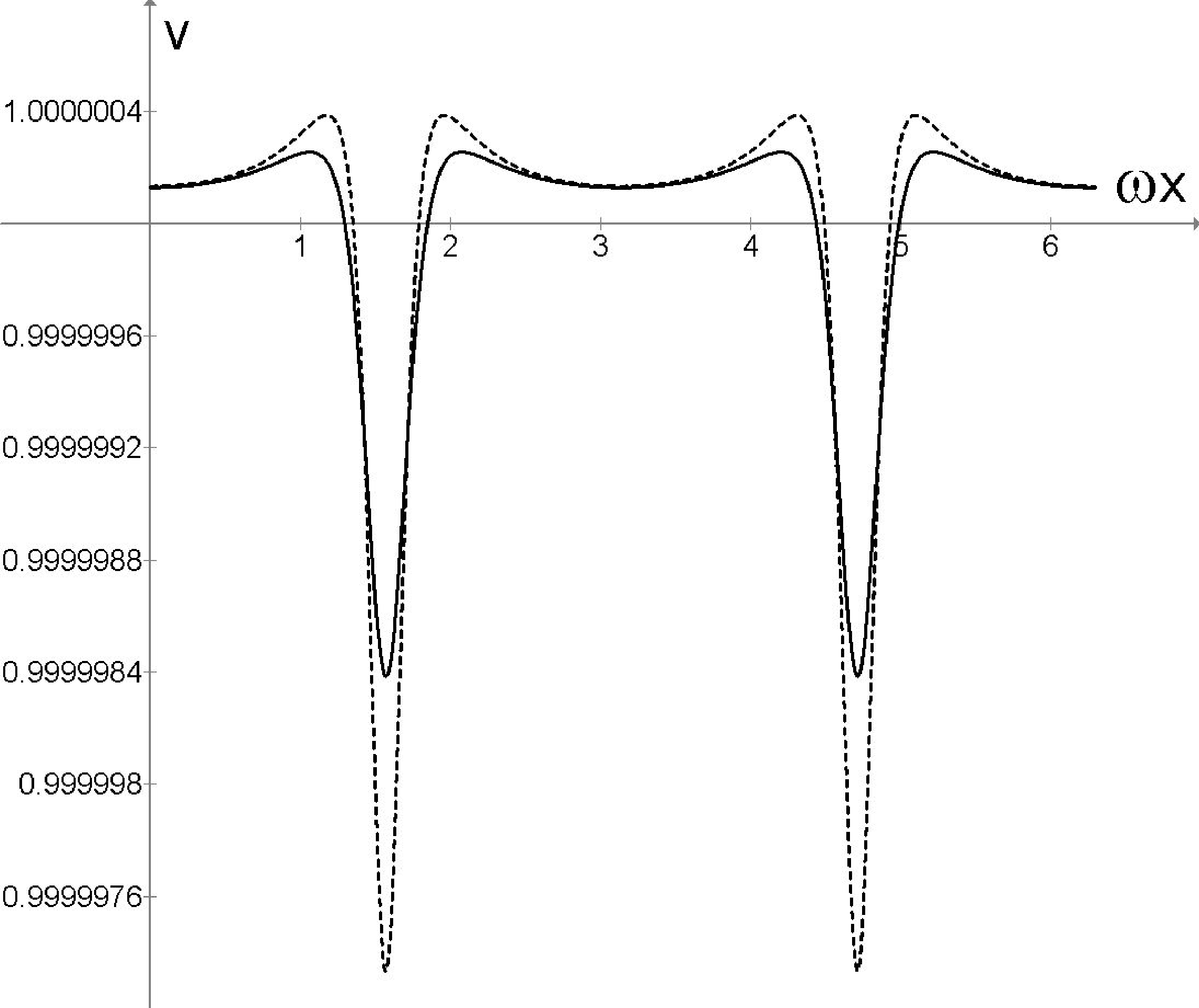}}
\caption{Oscillations of neutrino velocity.
The two cases are defined by $\sin^2 2\theta = 0.92$
(solid line), and $\sin^2 2\theta = 0.95$ (dashed line).} \label{fg2}
}
%
see, most of the time it exceeds the speed of light. In particular, $v(x) > 1$ for
all $x \ll L$. We see that averaging $v(x)$ over small distances results in
superluminal effective speed. The effective speed $v_{eff}$ is defined as
\begin{equation}\label{veff}
v_{eff} \equiv \frac{x(t)}{t} \approx 1- \frac{\tau(x)}{x} \,.
\end{equation}
As opposed to $v$, that can not be directly measured, the effective speed
$v_{eff}$ can. In fact, it is $v_{eff}$ that has been measured in the
experiments \cite{a,b,1}. Using (\ref{477}) in (\ref{veff}), we easily calculate the
formula for the effective speed. Its graph is displayed in Fig. \ref{fg3}.
%
\FIGURE{
\hbox{\epsfysize=7.2cm\epsffile{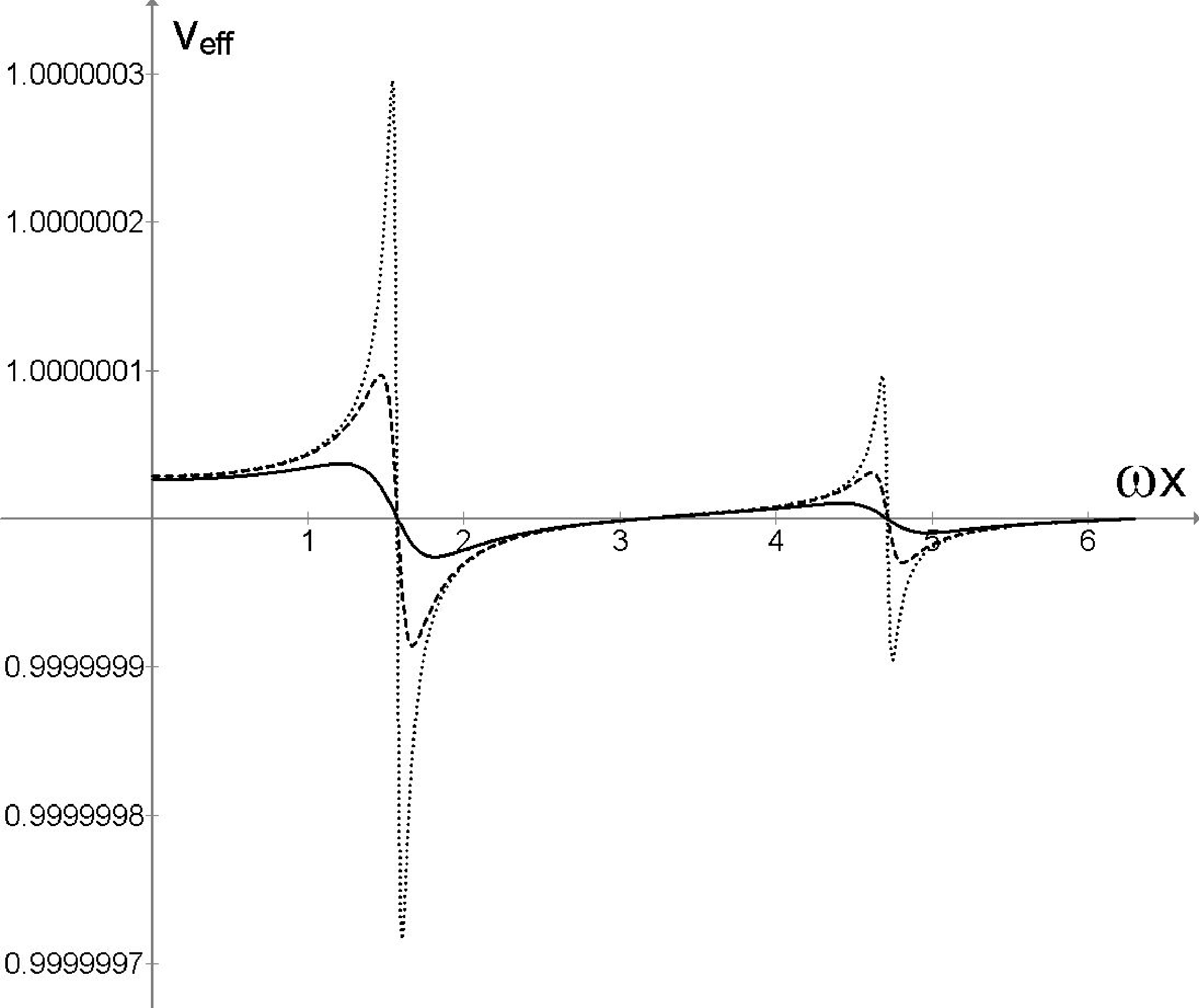}}
\caption{Oscillations of neutrino effective velocity.
The three cases are defined by $\sin^2 2\theta = 0.92$
(solid line), $\sin^2 2\theta = 0.99$ (dashed line), and $\sin^2 2\theta = 0.999$
(dotted line).} \label{fg3}
}
%
As we can see, it is not exactly a periodic function, but nevertheless, it
periodically exceeds the speed of light.

Let us now calculate the extreme values of neutrino delays and velocities.
To simplify exposition, we shall restrict our analysis to one period $x \in [0,L]$.
Then, if the extreme points are denoted by $x_{\pm}$ ($\tau_{min} \equiv
\tau(x_{-})$, $\tau_{max} \equiv \tau(x_{+})$), one finds $x_+ + x_- = L$,
and $\tau_{max} = - \tau_{min}$. The extreme points $x_+$ and $x_-$ are
obtained by solving the equation $d\tau/dx = 0$, whereupon
$$
x_{\pm} \approx \frac{L}{2} \left[1 \pm \left(1-
\frac{4 \theta}{\pi}\right)\right] \,,
$$
and the function $\tau_{min} \equiv \tau(x_{-})$, in the linear approximation
in $\tau/\ell$, takes the value
$$
\tau_{min}  = -  \,  \frac{\omega \ell^2}{4 }  \,
\frac{\sin^2 2 \theta}{\cos 2 \theta } \,.
$$
The corresponding effective velocity is given by $v_{eff}(x_-) =
1-\tau_{min}/x_-$. To illustrate this, let us calculate the minimal value
that $\tau(x)$ can possibly have for a given $\theta$, and display the
corresponding $x_-$ and $v_{eff}$. Fixing our free parameters as in the
example (\ref{example}), we obtain the Table \ref{t1}.
As we can see, the minimal time neutrino needs to arrive at $x$ depends on
the mixing angle $\theta$. So does the corresponding effective velocity
$v_{eff}$. The maximal effect is obtained if $\theta$ is close to $\pi/4$.
\TABLE{
\caption{$\theta$ dependence of the minimal neutrino delay} \label{t1}
\begin{tabular}{|c|c|c|c|}
\hline
$\sin^2 2\theta$ &    $x_{-}$        &  $\tau_{min}$       &  $v_{eff}$                        \\ \hline
0.92                  &  $7,400$ km    & $-1\, {\rm ns}$     &  $1+1.48 \cdot 10^{-7}$   \\
0.99                  &  $8,400$ km    & $-3\, {\rm ns}$     &  $1+4.22\cdot 10^{-7}$    \\
0.999                &  $8,820$ km    & $-10\, {\rm ns}$   &  $1+13.5\cdot 10^{-7}$     \\
\s 0.9999\s   &\s $8,940$ km\s &\s $-32\, {\rm ns}$\s &\s $1+42.2\cdot 10^{-7}$\s \\
\hline
\end{tabular}
}
Interestingly enough, the corresponding $x$ turns out to be close to the
value $x = L/2$ for all the displayed $\theta$. Unfortunately, the
probability density $P_{\mu}$ in $x = L/2$ goes to zero when
$\theta\to\pi/4$. Indeed, it is seen from (\ref{17d}) that complete
disappearance of muon neutrino is possible only if $\theta =\pi/4$.
Whenever this happens, a pure tau neutrino appears instead.

Similarly, the maximal value of the velocity $v(x)$ is obtained by solving
the equation $dv/dx =0$. It is shown that $v_{max}$ increases with
$\theta$, approaching its absolute maximum $v_{max} = 1+ \frac{2}{3}\,$
when $\theta \to \pi/4$. At the same time, $x \to L/2$. Sadly, the
probability $P_{\mu}$ to detect muon neutrino in the point where it reaches
its maximal velocity is zero. In other words, {\it while
accelerating towards $v_{max} = 1+ \frac{2}{3}$, the muon neutrino
gradually disappears}.

As for the maximum of the effective velocity $v_{eff}$, it is obtained as
the solution of the equation $dv_{eff}/dx = 0$. Again, the maximum of
$v_{eff}$ is a function of $\theta$, which approaches its absolute maximum
$(v_{eff})_{max} = 1 + 2\ell/L$ in the limit $\theta \to \pi/4$. At the
same time, $x \to L/2$, and the probability density $P_{\mu}$ goes to zero.

Finally, let us mention that the case $\theta = \pi/4$ is completely
different from the case $\theta \to \pi/4$. Indeed, when $\theta = \pi/4$,
the velocity $v(x)$ always exceeds the speed of light. Its behaviour is
shown in Fig. \ref{fg4}. As we can see, it reaches its maximal value
%
\FIGURE{
\hbox{\epsfysize=7.2cm\epsffile{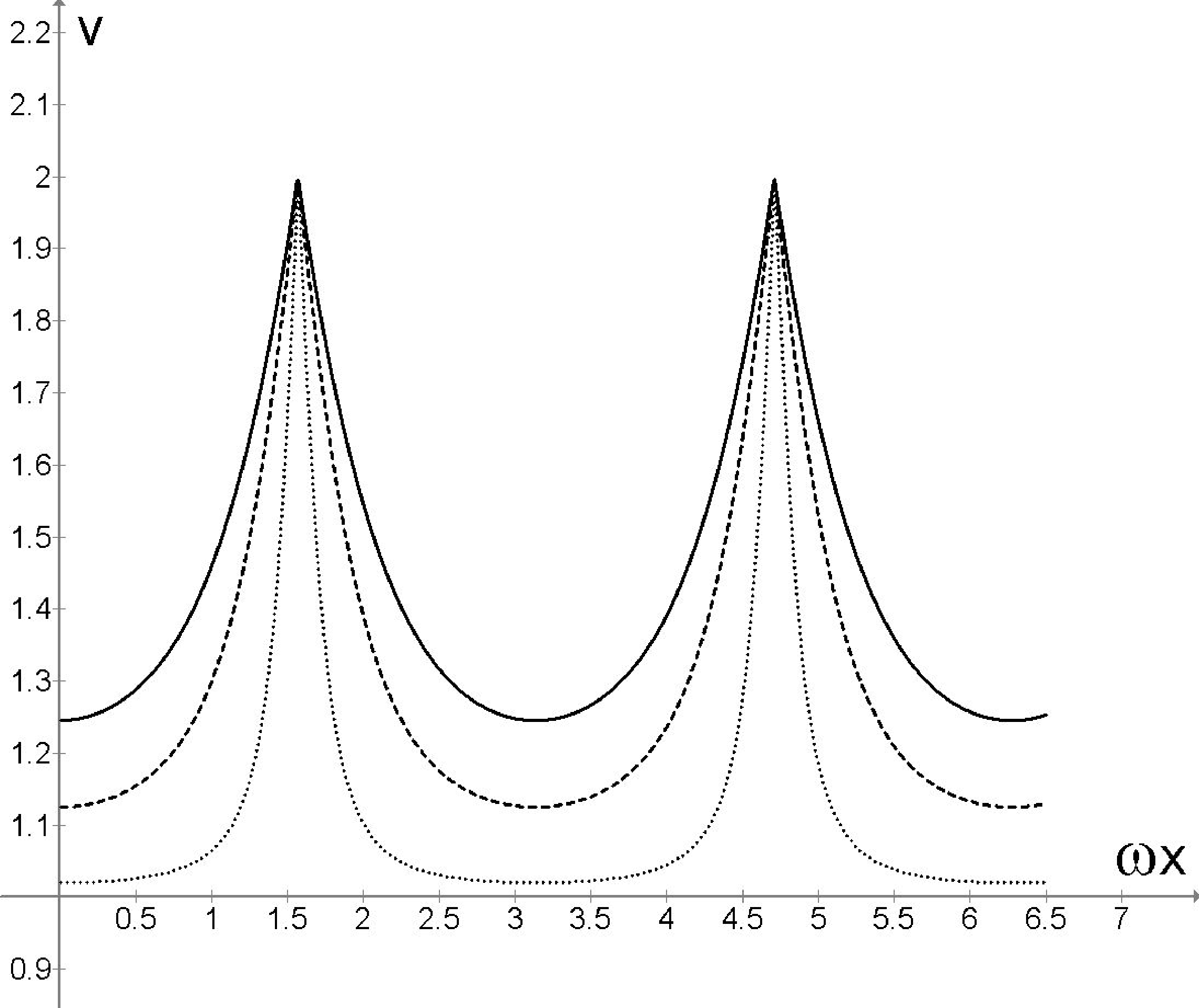}}
\caption{Oscillations of neutrino velocity for $\theta = \pi/4$.
The three cases are defined by $\omega\ell = 0.7$
(solid line), $\omega\ell = 0.5$ (dashed line), and $\omega\ell = 0.2$
(dotted line).} \label{fg4}
}
%
$v_{max} = 2$ in $x = L/2$, independently of the neutrino energy.
Unfortunately, the probability density to detect $v_{max} = 2$ turns out to
be zero. Again, this is because the muon neutrino turns into tau neutrino
in $x = L/2$.

\section{\label{shape}The shape of the wave packet}

Let us analyze how our results depend on the shape of the wave packet. It
is useful to introduce some parameter $\gamma \geq 0$, so that variation of
$\gamma$ from zero to infinity changes the shape of the wave packet from
almost rectangular to very sharp. Let us generalize (\ref{23}) by choosing
\begin{equation}\label{41}
\rho (\tau) \propto \left\{
\begin{array}{cl}
\ds \left(1- \frac{\tau^2}{\ell^2}\right)^\gamma , & -\ell < \tau < \ell   \\
0 \,, & \ \,{\rm otherwise}\ .
\end{array} \right.
\end{equation}
To check if the corresponding momentum distribution is sharp around $p=p_0$,
as required by our introductory assumptions, we calculate the Fourier
transform
\begin{equation}\label{42}
a(p) = \frac{1}{\sqrt{2\pi}} \int d \tau \, \rho (\tau)\,
e^{i(p-p_0)\tau}   \,.
\end{equation}
The resulting expression has the form
\begin{equation}\label{43d}
a(p_0 + p) \propto \frac{J_{\gamma+\frac{1}{2}}(\ell p)}
{(\ell p)^{\gamma+\frac{1}{2}}}  \,,
\end{equation}
where $J_{\gamma+\frac{1}{2}}(p)$ are ordinary Bessel functions of the order
$\gamma+\frac{1}{2}$. Note that $a(p)$ is real and finite function of its
argument. For integer values of $\gamma$, it can be expressed in terms of
elementary functions. For example, the expression for $\gamma=0$ reads
\begin{equation}\label{43e}
a(p_0+p) \propto \frac{\sin \ell p}{\ell p} \, ,
\end{equation}
and the expression for $\gamma=1$ has the form
\begin{equation}\label{43c}
a(p_0+p) \propto  \frac{\sin \ell p}{(\ell p)^3}(1- \ell p \cot \ell p) \, .
\end{equation}
The necessary and sufficient condition for both amplitudes to be
well localized around $p=p_0$ can be expressed in terms of the
dimensionless quantity $\ell p_0$. It reads
\begin{equation}\label{con}
\ell p_0 \gg 1 \, .
\end{equation}
Indeed, in the limit $\ell p_0 \to \infty$, the distributions
(\ref{43e}) and (\ref{43c}) take the form $a(p) \propto \delta
(p-p_0)$. As an illustration, the data from the recent
experiments \cite{a,b,1} yield $\ell p_0 > 10^{14} \gg 1$,
telling us that the condition (\ref{con}) is indeed satisfied.
In conclusion, the momentum distributions for the whole interval
$0 < \gamma < 1$ are sharply localized.

Now that we are convinced that the approximation we have chosen
to work with holds true, we proceed to solve the equation
(\ref{19}) for the set of parameter dependent amplitudes
(\ref{41}). Neglecting terms proportional to the small quantity
$\btau_2$, the equation (\ref{19}) turns into the quadratic
equation
$$
{\btau_1}\tau^2 + \gamma\ell^2\,\tau - {\btau_1}\ell^2 = 0 \,,
$$
which differs from the corresponding equation of the case $\gamma = 1$ by
the simple replacement $\btau_1 \to \btau_1/\gamma$. Having this in mind,
we easily obtain
\begin{equation}\label{47}
\tau (x) = -\frac{\gamma \ell^2}{2 {\btau_1}}\left(1-
\sqrt{1 + \left(\frac{2 \, {\btau_1}}{\gamma\ell}\right)^2} \right) \,.
\end{equation}
In the lowest approximation in $\tau/\ell$, we come to
\begin{equation}\label{47a}
\tau (x)= \frac{{\btau_1}}{\gamma} \,.
\end{equation}
The requirement that (\ref{47a}) has a dominant role in
(\ref{47}) puts a restriction on $\gamma$. For the experiments
\cite{a,b,1}, a sufficient condition to ensure the validity of
(\ref{47a}) is
$$
\gamma \gg 10^{-8}.
$$

Finally, let us say something about how neutrino velocity
depends on $\gamma$. We have seen in the preceding section that
$v_{max}$ is a discontinuous function in $\theta = \pi/4$. The
same happens in the case $\gamma \neq 1$. Indeed, the neutrino
velocity has two absolute maximums,
\begin{eqnarray}
 v_{max} = 1 + \frac{2\gamma}{3} & \quad {\rm when} \quad  &
      \theta\to\frac{\pi}{4} \,,         \nonumber  \\
 v_{max} = 1 + \gamma & \quad {\rm when} \quad &
      \theta =\frac{\pi}{4}\,,            \nonumber
\end{eqnarray}
which both take place in $x = L/2$. In this point, however, the
probability density $P_{\mu}$ goes to zero. Thus, we can say
that {\it muon neutrino, which approaches the point of
absolutely maximal velocity, gradually disappears}. This is
because it turns into a tau neutrino.
\FIGURE{
\hbox{\epsfysize=7.2cm\epsffile{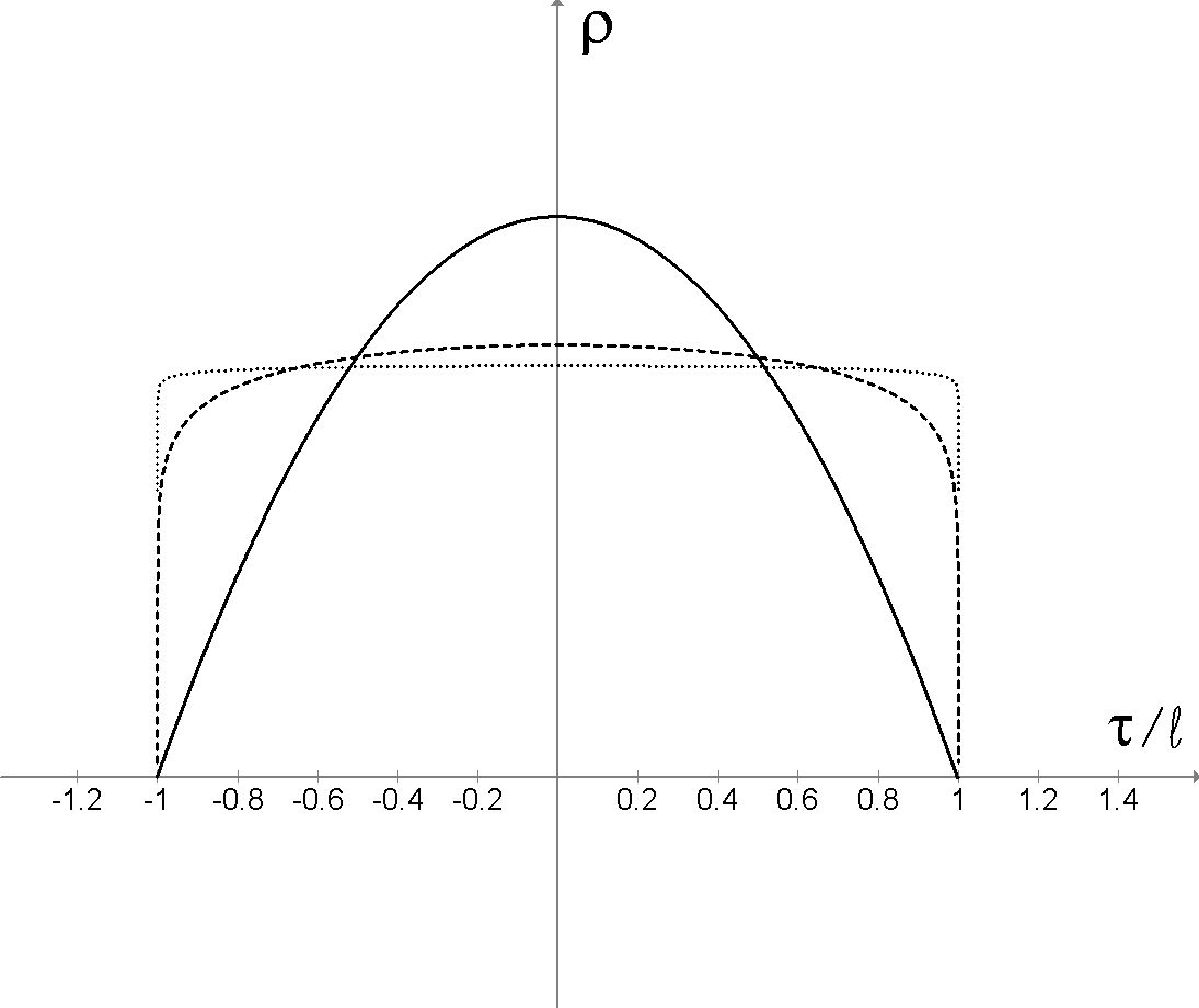}}
\caption{Shapes of the neutrino wave packet. The three
cases are defined by $\gamma = 1$ (solid line), $\gamma = 0.1$ (dashed line),
and $\gamma = 0.01$ (dotted line).} \label{fg5}
}

In conclusion, as $\gamma$ approaches zero, the wave packet is
more sharply localized in momentum space, and has more
rectangular shape in coordinate space. (See the example in Fig.
\ref{fg5}.) This makes the neutrino delay time $|\tau|$
increase. As a consequence, the superluminal effect becomes
easier to detect. Still, as the known experiments lack the
parameter that determines the shape of the neutrino wave packet,
the comparison with measurements remains an uneasy problem.

At the end, let us note that the above results refer to the
detection of muon neutrino. The generalization to the case of
tau neutrinos is straightforward. In the next section, we shall
analyze the evolution of tau component of initially pure muon
neutrino.

\section{\label{tau}Velocity oscillations of tau neutrinos}

In the preceding section, we have considered the dynamics of
initially pure muon neutrino, and calculated the probability to
detect muon neutrino at later times. In a similar way, we can
derive the probability density to detect tau neutrino. It is
defined by
$$
P_{\tau}(x,t)=\left|\langle\nu_{\tau}(x)|\nu_{\mu}(t)\rangle \right|^2 \,,
$$
where $|\nu_{\tau}(x)\rangle \equiv \sin\theta\, | 1 \rangle | x \rangle +
\cos\theta\, | 2 \rangle | x \rangle$ is the eigenstate of the position
operator for tau neutrino. The direct calculation yields
\begin{equation}\label{18}
P_{\tau}(x,t) = \rho^2 (\bar v\tau)\sin^2 2\theta\,\sin^2 \omega t \,.
\end{equation}
As we can see, no correction linear in $\Delta v$ appears in (\ref{18}).
It is checked that
$$
\int\left(P_{\mu}+P_{\tau}\right)dx = \int \rho^2(x) dx = 1 \,,
$$
as it should be in the approximation we work with.

If the neutrino detector detects tau neutrinos instead of muon
neutrinos, the corresponding equation $\partial P_\tau / \partial
t = 0$ takes the form
\begin{equation}\label{19d}
\frac{\partial_{\tau}\rho^2}{\rho^2} + \frac{4}{\ell^2}
\btau_3(x) =0   \, ,
\end{equation}
where
\begin{equation}\label{40}
\btau_3(x) \equiv \frac{\omega \ell^2}{4}  \cot\omega x  \,.
\end{equation}
As we can see, the dependence on the mixing angle $\theta$ does
not appear in (\ref{40}). Choosing the same form for $\rho$ as
in (\ref{41}), we obtain the expression of the same form as
(\ref{47}), with the only difference that $\btau_3(x)$ is used
instead of $\btau_1(x)$. The approximation linear in $\tau/\ell$
then yields
\begin{equation}\label{40a}
\tau_{\tau}(x) = \frac{\btau_3(x)}{\gamma} \,.
\end{equation}

If we restrict our analysis to distances which are not too close
to $(n+\frac{1}{2})L$, and the mixing angle $\theta$ is close to
$\pi/4$, the corresponding formula for $\btau_1$ takes the
simplified form $\btau_1 \approx - \frac{\omega \ell^2}{2} \tan
\omega x$. It then yields
\begin{equation}\label{40b}
\btau_1(x)\, \btau_3(x) \approx -
\frac{\omega^2 \ell^4}{8} = {\rm const.} \, ,
\end{equation}
or equivalently,
\begin{equation}\label{40c}
\tau_\mu (x)\,  \tau_\tau (x) \approx - \frac{\omega^2 }{8\gamma^2}
\left[\ell^2 - \tau^2_\mu (x) \right] \left[\ell^2 - \tau^2_\tau (x) \right]\,.
\end{equation}
As we can see, the arrival times $\tau_\mu$ and $\tau_\tau$ for muon and
tau neutrinos have opposite signs. Thus, when one of them has superluminal
speed, the speed of the other is subluminal. The relation (\ref{40c})
nicely illustrates connection between oscillations of muon and tau
neutrinos. When one flavor has minimal deviation ($\tau_\mu (x) \to 0$) the
other has maximal one ($\tau_\tau (x) \to \ell$). (Note, however, that our
results are derived under the assumption that $\tau$ is not too close
either to $\ell$ or to zero.)

Finally, let us point out another interesting relation between
$\btau_1(x)$ and $\btau_3(x)$:
$$
\btau_1 \left(x + L/2\right) \approx 2\, \btau_3 (x) \,.
$$
With this, the calculations related to tau neutrino are greatly
simplified. In particular, the linear approximation in
$\tau/\ell$ yields the relation $\tau_\mu \left(x + L/2\right) =
2\, \tau_\tau (x)$, showing that the double delay time of tau
neutrino is obtained by the simple shifting $x \to x + L/2$ in
the expression for the delay time of muon neutrino.

\section{\label{summary}Summary and discussion}

In this paper, we have demonstrated that the flight of a free neutrino is
characterized not only by the well known flavor oscillations, but also by
the oscillations of neutrino velocity. This has been done by considering
the free evolution of initially pure muon neutrino. First, we calculated
the probability density for detecting muon neutrino by the detector placed
in a fixed position $x$. Such a probability distribution was a function of
time, and its maximum was identified with the moment neutrino arrived at
$x$. This way, we obtained the formula for the evaluation of time the
neutrino needed to fly across the distance $x$.

It should be noted that our formula differs from similar formulae found in
literature (see, for example, \cite{2,3,4}). The reason for this is
the difference in our definitions of the neutrino position. While we
identify it with the maximum of its probability distribution, the authors of
\cite{2,3,4} use the average of neutrino position operator. As a
consequence, our formula carries additional dependence on the size and
shape of the neutrino wave packet. The neutrino delay, as compared to
the arrival time of the photon, was found to be a periodic function of $x$,
with the period which did not depend on the size and shape of the neutrino
wave packet. The neutrino velocity, on the other hand, turned out to
drastically depend on the size and shape of the wave packet: the more
rectangular the shape of the packet was, the higher velocity neutrino could
reach. As it turned out, the {\it neutrino velocity periodically exceeded
the speed of light}. We were able to derive the maximal velocity the
neutrino could possibly achieve during its flight. As it turned out, the
probability to detect maximal velocity was rather small, approaching zero
when $\theta \to \pi/4$.

The fact that our formula depends on the size and shape of the neutrino
wave packet makes the comparison with the experiment very difficult. This
is because the known experiments do not provide the information on these
two parameters. Nevertheless, we shall try to test our formula by comparing
its predictions with three recent experiments \cite{a,b,1}. To this end, we
shall estimate the size of the neutrino wave packet using the results found
in literature. In particular, a useful study of the properties of
accelerator neutrinos can be found in Ref. \cite{c}. Using the size of the
wave packet as found in \cite{c}, and almost Gaussian shape as defined by
$\gamma = 1$, we make the Table \ref{t2}.
\TABLE{
\caption{Comparison with experiments}\label{t2}
\begin{tabular}{|c|c|c|c|c|c|}\hline
Experiment &  $p_0$  &  2$\ell$  &  $x$  &  $v_{exp}-1$  &  $v_{eff}-1$                 \\ \hline
MINOS & $3$ GeV & $7.7$ cm & $734$ km & $(5.1\pm 7.5)\cdot 10^{-5}$ & $0.9\cdot 10^{-15}$  \\
ICARUS & $17$ GeV & $0.67$ cm & $730$ km & $(0.1\pm 5.7)\cdot 10^{-6}$ & $1.5\cdot 10^{-19}$ \\
OPERA &\q $17$ GeV \q&\q $0.67$ cm\q &\q $730$ km\q &\q $(2.7\pm 6.5)\cdot 10^{-6}$\q &\q $1.5\cdot 10^{-19}$\q \\
\hline
\end{tabular}
}
As we can see, our theoretical predictions do not contradict any of the
three experiments. Even if the neutrino wave packet is taken $10^5$ times
longer, the agreement with the experiment is not compromised. The same
holds when it comes to the wave packet shape. Indeed, the change of
$\gamma$ in the allowed interval $\gamma \gg 10^{-8}$ leaves the
theoretical values within the experimental error. In fact, it is the big
experimental error that basically ensures this agreement. For a real test
of our theoretical predictions, more efficient measurements are needed.
In particular, our equations suggest how neutrino free parameters should
be chosen to maximize the superluminal effect.

Let us say something about energy dependence of the neutrino
velocity. As seen from our formulae, the energy dependence of
$v_{eff}-1$ has oscillating character. To simplify the analysis,
we shall restrict to the region $x \ll L$, which is achieved by
using short range high-energy neutrinos. Upon this, the measured
quantity $v_{eff}-1$ becomes proportional to $\ell^2/p_0^2$,
which reduces to $\ell^2/E_0^2$ in the ultrarelativistic limit.
At the same time, the wave packet size $\ell$ is also energy
dependent. Indeed, it has been shown in \cite{c} that, depending
on the experimental details, $\ell$ is proportional to either
$1/E_0$ or $1/E_0^2$. Thus, in the ultrarelativistic limit,
$$
v_{eff}-1\, \sim\, \frac{1}{E_0^4} \qquad {\rm or} \qquad
v_{eff}-1\, \sim\, \frac{1}{E_0^6} \,.
$$
This is a strong energy dependence, but still undetectable by
the ICARUS and OPERA experiments. (MINOS experiment is excluded
from the analysis because it violates the requirement $x \ll
L$.) Indeed, it is seen from Table \ref{t2} that the values of
$v_{eff}-1$ are $10^{13}$ times smaller than $v_{exp}-1$, which
implies that {\it $100$ times lower energy still yields the
result that agrees with the experiments}. Higher energies, on
the other hand, diminish the value of $v_{eff}-1$, and
therefore, fit the experiments even better.

Finally, let us comment on the result of Ref. \cite{d} stating that
superluminal neutrinos should rapidly loose energy during their flight.
This result has been derived with the assumption that neutrino speed does
not change along neutrino trajectory. However, we have shown that this is
not the case. In general, the neutrino velocity has oscillating character,
but we shall simplify the analysis by adopting the condition $x \ll L$. In
this regime, the results of the preceding sections yield
$$
\delta\, \approx\, 2\alpha \frac{d}{dx} \Big( \frac{x}{E^2} \Big)\,,\qquad
\alpha \equiv \frac{1}{2\gamma} \bigg(\frac{\Delta m^2 \ell}{4}\bigg)^2 ,
$$
where $\delta \equiv v^2-1$. Using this expression in the formula for
the rate at which superluminal neutrinos loose their energy \cite{d},
$$
\frac{dE}{dx} = - \kappa E^6 \delta^3 \,, \qquad
\kappa \equiv \frac{25}{448}\,\frac{G_F^2}{192\pi^3}\,,
$$
we obtain a complicated, higher order differential equation. To simplify
the calculations, we shall assume that superluminal neutrinos loose their
energy slowly ($|dE/dx| \ll E/x$). Then, we obtain a simple approximative
solution
$$
E \approx E_0 - E_T \,,\qquad  E_T \equiv 8\kappa\alpha^3\,x \,,
$$
telling us that the rate superluminal neutrinos loose their energy is
linear in $x$. Notice, however, that the value of the constant
$\kappa\alpha^3$ is extremely small in the present experiments. For the
numerical illustration, we shall take data from Table \ref{t2}. With the
known value of the Fermi coupling constant $G_F \approx 1.17 \cdot
10^{-5}\,{\rm GeV}^{-2}$, and using the almost Gaussian wave packet as
defined by $\gamma = 1$, we find
$$
E_T\, \lesssim\, 1.1 \cdot 10^{-35}\ {\rm GeV} \,.
$$
Having in mind that the initial neutrino energy in the considered
experiments \cite{a,b,1} is $E_0 \geq 3$ GeV, we see that the loss of
energy during the flight of superluminal neutrinos is negligible. Our
initial assumption $|dE/dx| \ll E/x$ is thereby justified, and the validity
of our conclusion is confirmed. We can still change the wave packet shape,
but this can not significantly modify our conclusions. Indeed, the lowest
allowed value of $\gamma$ is of the order $10^{-7}$, which leads to the
value $E_T \sim 10^{-14}$ GeV. This is still negligible with respect to the
initial neutrino energy ($E_T \ll E_0$) in all terestrial experiments.
Summarized, {\it we have proven the existence of energy conserving
superluminal free neutrinos}.

In conclusion, if neutrinos have different masses, the oscillations of
neutrino velocity necessarily exist. In particular, the neutrino velocity
periodically exceeds the speed of light. The significance of this result is
threefold. First, it shows that superluminal speed can be achieved without
violation of special relativity. Second, our equations suggest how neutrino
parameters should be chosen to maximize the superluminal effect in new
experiments. Finally, our formula offers an independent way to determine
neutrino mixing angles.

\acknowledgments
This work is supported in part by the Serbian Ministry of Education and Science,
under Contract No. $171031$.

\end{document}